\documentclass[conference]{IEEEtran}
\usepackage{hyperref}
\usepackage{breakurl}

\begin{document}

\title{Benchmarking Obfuscators of Functionality}
\author{Clark Thomborson\\
\url{cthombor@cs.auckland.ac.nz}\\
University of Auckland\\
\\
Version 1.0 of 13 January 2015\\
Submitted to SPRO 2015, \url{https://aspire-fp7.eu/spro/}\\
Available on arXiv}

\maketitle
\thispagestyle{plain}
\pagestyle{plain}

\begin{abstract}
  We propose a set of benchmarks for evaluating the practicality of
  software obfuscators which rely on provably-secure methods for
  functional obfuscation.\par \emph{Note to SPRO referees:} this paper
  is one page longer than the 7-page limit for a regular submission.
  I will prepare a 7-page version, if this is required for publication.
\end{abstract}


\section{Introduction}

Recent advances in cryptographic theory have pointed the way toward
constructions of provably-secure \emph{indistinguishability
  obfuscators} for Boolean functions \cite{Garg2013}.  However, as
with many other theoretical advances, the reduction to practice may be
problematic.  The constructions may be very difficult to implement;
the constructions may ``leak'' information through side-channels
that are not considered by the theoretical proofs; and the obfuscated
functions may be ``bloated'' to the point that they are not feasibly
computable on a handheld device, a desktop computer, or even on
a supercomputer.

This article is an early response to the 30 September 2014
announcement of the SafeWare research program, managed by the US
Defense Advanced Research Projects Agency (DARPA) \cite{SafeWare2014},
which will explore the practical feasibility of provably-secure
obfuscation, as well as to advance the theory of such obfuscations.

Any obfuscators which are constructed under SafeWare will be evaluated
for their runtime overhead (average and worst-case), their obfuscation
security level (\emph{e.g.} an adversary work factor), and any
potential side-channel vulnerabilities.  In this article we propose a
framework for evaluating runtime overheads.

SafeWare-funded researchers will attempt to construct obfuscated
programs which are provably secure, \emph{i.e.} programs whose
de-obfuscation would involve the solution of a problem which is known,
or generally believed, to be computationally infeasible.  Currently,
the only plausible candidates for such constructions are what we would
call functional obfuscators, as opposed to control-flow obfuscators,
data obfuscators, system-call obfuscators, communication obfuscators,
or semantic obfuscators (in which portions of the program are
expressed in a programming language that is initially unknown to the
adversary).

Current techniques in control-flow flattening, opaque predicates, and
the breaking of abstractions \cite{Collberg2002} are specifically
excluded from attention in SafeWare, because programs obfuscated by
these techniques may be de-obfuscated without solving a
computationally-hard problem.  It remains an open question, to
be addressed by SafeWare-funded theoreticians, whether or how
the control-flow graph of a program could be encoded into an obfuscated
function which does not leak important information about control
flow to a reverse engineer who performs a dynamic analysis on
the obfuscated program.

Our focus in this article is on what we call functionally-obscure
programs.  We say a program is functionally-obscure if it contains an
obfuscated function whose behaviour is required for program
correctness, \emph{i.e.} if a change to its value at any point might
cause the program to behave incorrectly.  Functional obscurity is not,
in and of itself, a solution to digital rights management of software,
because an attacker may replace an obfuscated password-recognition
function by a stub which returns true for any input, or they may
invert the comparison logic so that the program accepts any password
except the correct one \cite{LaDue1997}.  However functional obscurity
may still be an important line of defense in a digital rights
management system, and it may also be used to meet other security
goals.  In particular, functional obscurity would significantly
mitigate the risk of password leakage, if password-recognition
functions are securely obfuscated in a computational environment which
is well-secured against adversarial observation and control.

The most promising line of research into functionally-obscure
obfuscators is, at present, based on the security model of
``indistinguishability obfuscation'' with respect to a set of
circuits, such as $\mbox{\rm NC}^1$.  This set of functions may be,
informally, considered to be a ``crowd'' of functions within which any
individual function of practical interest would be ``anonymous''
(\emph{i.e.} indistinguishable from any other member of the crowd) --
even after an attacker has spent a long time probing its behaviour (by
observing its output on adversarially-controlled inputs) in an effort
to determine its secret identity.

Indistinguishability obfuscation is an appropriate security concept
for password-recognition functions, and for all other ``point''
functions (such as signature-verifiers).  An attacker who is unable to
deobfuscate, or to exploit a side-channel, must perform an exhaustive
search over all likely inputs to the obfuscated function to discover
the point at which its output changes.  If a securely-obfuscated
password-recognition function can be efficiently evaluated on low-cost
computing platforms, and if it can be economically produced by an
obfuscation process, this would be of great practical utility.
However if the obfuscation is weak, then the obfuscation is a
dangerous waste of computational resource.  This line of reasoning
suggests that the provision of adequate security is the primary
requirement on a functional obfuscator.

In this article, we do not attempt to evaluate the adequacy of
security -- this is a very challenging technical problem which
includes the construction of a valid security model.  Instead, we
focus on the easier, but still quite challenging and important,
problem of evaluating the performance of an obfuscated function.

The runtime performance of a securely-obfuscated function is always a
satisficing requirement for its end-users: rapidly-evaluating
functions are preferable to slowly-evaluating functions.  However, in
any given application, the maximally-acceptable runtime performance is
a feasibility constraint.  This constraint may be extremely
challenging, or even infeasible, to satisfy.  For example, in the
specific case of password-recognition functions, end-users will not
wait years, and they may not even be willing to wait seconds, for a
program to accept or reject their password.

We cannot predict the most important applications of obfuscated
function, so we cannot benchmark a general technique for secure
obfuscation against a fixed-time threshold.  However we can establish
some indicative runtime constraints, \emph{e.g.}  we might insist that
the recognition of an 8-character password must be accomplished within
1 second on a mid-spec smartphone such as a Samsung Galaxy S III.
Highly-specific performance requirements of this nature are very
important in acceptance-testing, but they would provide little or no
guidance to theoreticians whose insights and theorems are based on
asysmptotic analysis..

It is technically challenging -- and this is the primary technical
focus of our article -- to construct an easily-assessed measure of
runtime performance which is valid, at least as a rough approximation,
on a wide variety of contemporary computing platforms, for a wide
variety of functions which might plausibly be obfuscated.

In Section 2, we argue that the runtime cost of functional obfuscators
should be estimated as $n\sqrt{w}$, where $n$ is the number of 2- or
3-input gates in the obfuscated circuit, and $w$ is the width of a
program which represents the obfuscated circuit.  We believe that this
functional form is simple enough to guide asymptotic analyses, while
being accurate enough to provide appropriate guidance.

Our definition of the width of a program is a significant restriction
on the usual notion of circuit width, because (in version 1 of our
file format) a program of declared width $w$ which represents a
Boolean circuit may have at most $w^2$ Boolean inputs and at most
$w^2$ Boolean outputs.  In subsequent versions of our file format,
after experimentation on the range of contemporary computational
platforms and functions of practical interest, we may relax this
restriction, perhaps allowing as many as $w^4$ inputs and outputs
(with a significant time-penalty for such extended-IO) to a width-$w$
program.

We have not parameterised our cost function on the depth of the
circuit, even though such cost functions have been researched
extensively in circuit complexity theory, because we are benchmarking
low-cost computational platforms which are evaluating Boolean
functions with billions or trillions of logic gates.  Our cost
estimate is intended to model the effects of the memory (or I/O)
bottleneck that will arise when the platform's evaluation of a Boolean
function requires a working set which exceeds the cache (or
main-memory) capacity of the platform. Circuit depth would only become
important if it were impossible to avoid a CPU bottleneck, \emph{i.e.}
in the case of very narrow circuits.  We do not expect this case to
arise in practical applications of obfuscated functions, for the
reasons discussed in Section 2.

In Section 3, we propose a space-efficient and
computationally-appropriate file format (BPW) for the evaluation of
very large Boolean functions with bounded program width.  There are
many existing formats for describing Boolean circuits, and any of
these might be used for describing obfuscated functions.  Some formats
are restricted to combinational logic, and therefore may be more
compact than formats which support sequential logic or those which
specify implementations such as programmable logic arrays or
full-custom integrated circuitry.  Some formats are designed to help
designers create attractive visual representations of small circuits.
We encourage future researchers on functionally-obscure software to
consider using any convenient representation when generating their
circuits, then translate into our representation when storing a very
large circuit in a computer file, or when evaluating a very large
circuit in software.  If we are funded to contribute to SafeWare, we
would envisage implementing routines to a translate
circuit-description files from our BPW format into a (very small
subset of) IEEE VHDL \cite{VHDL2008}, and vice versa.  By our
preference, and because SafeWare ``emphasizes the idea of creating and
leveraging open source architecture technology'' \cite{SafeWare2014},
our translation routines will be open-sourced.  We claim no
intellectual-property rights over the BPW format disclosed in this
article.

In Section 4, we propose an experimental method for evaluating our
proposed cost metric, to determine its range of validity for
contemporary computing platforms such as smartphones, laptops, and
desktop computers.

We summarise our contributions in Section 5.

\section{Cost Metric}

Obfuscated functions may be deployed occasionally on supercomputers,
however we believe most commmercially-important functionally-obscure
programs will be on mass-market platforms such as smart sensors, smart
phones, battery-powered laptops, and desktop computers.
Functionally-obscure programs could conceivably be used in
cloud-computing environments, and in ad-hoc distributed computing
environments, however the secure evaluation of a Boolean function in a
distributed environment has quite a different set of cost drivers due
to the latency and bandwidth limitations of communication links.
Readers who are interested in communication-limited functional
evaluations should review the literature on distributed secure
computations \cite{Bogetoft2009}, whereby geographically-separated
parties can provide secret inputs to a collaboratively-evaluated
function such as the result of an auction or a democratic vote.

The computational platforms of relevance to our context have limited
parallelism at any level of their memory hierarchy.  At any given
instant, there may be thousands or even tens of thousands of
register-level operations in progress; dozens of cache operations; 
a few main-memory operations; and one or two secondary-storage
operations.  Cache, main-memory, and secondary-storage operations
are of particular relevance, whenever a computer is evaluating a
Boolean function with millions or trillions of gate-equivalents,
unless the function is narrow enough that the working set of the
evaluation process will fit in the registers.

We base the analysis in the remainder of this section on the premise
that securely obfuscated programs must have widths of 50 or
more.

We define the term ``width of a program'' only informally in this
section, as a rough measure of its working set.  We will give this term a
formal definition in Section 3, when we define our BPW format.

Seven-character passwords, and (generally) cryptographic keys that are
shorter than 50 bits are susceptible to a brute-force attack; so we
use $w=50$ as the lower limit of our range of interest.  We encourage
SafeWare-funded security analysts to critically examine this lower
bound on $w$ for validity, that is, to determine whether or not there
is a general method of feasible attack on a width-$49$ BPW program.

The width of a circuit is well-established analytic concept
\cite{Pippenger1979, Dymond1989, Allender1990}.  If the gates of a
circuit are arranged in levels (or rows), such that each level has at
most $w$ gates, and such that the gate-outputs at each level are
connected only to gate-inputs on the next level, then the circuit has
width $w$.

The width of a function is the width of its narrowest gate-level
implementation in any circuit, in a given logic family (\emph{e.g.}
in 2-AND, 2-OR, and NOT gates).

A program which describes a circuit implementing a function may have a
width that greatly exceeds the width of the function.  We believe that
such unnecessarily-wide programs are the most promising candidates for
functional obfuscations, because the process of circuit analysis is
impeded very significantly by its width, and because it may be
very difficult, or even computationally infeasible, for an adversary
to discover a narrower implementation.

Wide circuits can be evaluated efficiently on parallel computers.  For
example, a circuit of width 50 or more may be evaluated by a 50-thread
computation of the following form.  Each thread runs a straight-line
program in which it fetches a few Boolean inputs, computes a simple
Boolean function, and writes a Boolean value into a shared memory
area.  It is not necessary to write programs with such explicit
parallelism, in order to exploit much of the parallelism available on
a modern CPU.  Hundreds or thousands of machine instructions may be
concurrently in the execution pipeline of a single CPU core, and a
single-threaded computation which relies on the instruction-scheduling
hardware is (in many cases) more efficient than a multithreaded
computation with explicit locks.  The CPUs on high-end
desktop computers may soon have some transactional memory features
\cite{Yoo2013} which could allow an extremely efficient multithreaded
evaluation of Boolean circuits -- if the evaluation state is small
enough to be held in L1 cache.

As indicated in the previous paragraph, modern computers have widely
differing numbers of CPU cores, and they have widely differing
organisations of their memory systems.  This diversity implies that a
circuit evaluator which is highly tuned for efficiency on one platform
may be very poorly optimised for another platform.  Our response to
this engineering challenge is to propose a special-purpose programming
language for the evaluation of large and wide Boolean circuits.  Our
language should be efficiently interpretable on any platform, and it
may be compiled with platform-specific optimisations if even
higher performance is required.  In the remainder of this section,
we identify the most important factors which affect the runtime
performance of a circuit evaluator on any platform, and we develop
some rough estimates of performance in particular cases, with
the goal of developing a general formula for predicting runtime
performance on any platform.

One of the key factors in any performance estimation is the size of
the working set of the program.  If the working set can be held
entirely in CPU registers, the computation will never be stalled on
cache accesses.  If the working set is cache-local, then the
computation will never be stalled on main memory accesses.
If the working set is small enough to fit in main memory, then
the computation will not thrash the secondary storage device.
Accordingly: our programming language has a wordsize of 1 bit,
so that its interpreter may (depending on the platform)
minimise the size of its working set by packing and unpacking
bits into machine words.  This is a CPU-memory tradeoff,
for the pack/unpack operations may result in a CPU-bottlenecked
evaluation which could be avoided (at the cost of occasional
cache faults) by storing Boolean values in machine bytes or
words rather than in machine bits.

As indicated earlier, we are restricting our attention to circuits
with width at least 50.  We are also restricting our attention to
circuits with at least millions ($10^6$) of 2- or 3-input gates.
Smaller circuits seem unlikely to be securely obfuscated.
Furthermore, initial constructions from an asymptotically-valid theory
such as indistinguishability obfuscation are rarely, if ever,
efficient with respect to constant factors and additive constants.

If a circuit with a million 3-input gates is \emph{not} organised for
temporal locality during its evaluation, then its width will be 500000
or more, and every gate-evaluation will require three fetches and one
store on a million-bit state vector.  This vector, even if it is
stored bitwise in 128 KB, is too large for L1-locality on most
contemporary computers.  However this state vector, and thus the
working-data set of the evaluation, will fit comfortably in the L2
cache of a smartphone, laptop, or desktop computer.  The evaluation of
each logic gate will thus involve four L1 misses: three reads and one
write.  A tightly-written inner loop for an interpreted evaluation
might execute thirty machine instructions per gate-evaluation, so we
estimate one L1 miss per eight machine instructions.  Modern
processors are very inefficient with such high L1 miss rates; their
memory systems are tuned for miss rates not exceeding a small fraction of a
percent, very roughly 1/300.  We conclude that a disorganised
evaluation will have a slowdown of a factor of approximately 300/8 = 40
in comparison to a memory-local (CPU-bottlenecked) evaluation of
a narrower circuit on the same platform.

High-end GPUs in desktop computers \emph{may} support very rapid
evaluations of Boolean circuits, so long as the circuit description is
compressed well enough to avoid a communication bottleneck at the
GPU-CPU interface.  We expect that a disorganised billion-gate circuit
would be evaluated by a high-end GPU at a rate approaching one
gate-evaluation per four DRAM cycles, that is, at (very roughly) 10
million gate-evaluations per second.  Note that, due to the
disorganisation, each a DRAM word in the 128 MB working set of this
billion-gate evaluation contains only a single bit of relevance to
its current stage of computation.

NVIDIA's Fermi architecture for its GPUs has 32k general-purpose
registers and 512 ALU cores \cite{Fermi2009}.  We thus expect the
working set of a well-optimised interpreter to be held in GPU
registers, if the Boolean circuit has width 10000 or less.  The
description of the billion-gate circuit will not be register-local,
but it could be streamed from main memory at a rate of perhaps 3 GB/s,
using DMA over its PCIe channel.  Our BPW format will encode gates in
very large circuits at (roughly) 8B/gate, so we estimate high-end GPU
performance on moderately-organised billion-gate Boolean circuits to
be 3/8 billion evaluations per second.  This is roughly 40 times
faster than our estimated performance for GPUs on disorganised
billion-gate circuits.

Based on the preceding performance estimates, we tentatively identify
$n$ (circuit size) and $w$ (program width) as the most important
factors controlling runtime performance on any platform, under the
constraints that $w \geq 50$ (so that the computation is at least
50-way parallelisable) and $10^6 \leq n \leq 10^9$ (so that it is
reasonable to assume the circuit description is in main memory,
thereby avoiding I/O bottlenecks).  If we aim only at predicting the
relative performance for two different evaluations on the same
platform, we need not encumber our predictions with platform-specific
parameters if we adopt a general model of memory system performance.
In prior work, we suggested one such model \cite{Thomborson1998}.
We will use this model to develop a performance estimate, immediately
after describing it briefly in the next paragraph.  In Section 4
we propose a set of experiments which would validate (or invalidate)
our performance model.

Our general model of memory system performance is based on the
assumption of a hierarchical memory system which may be visualised as
a triangle.  At the apex of the triangle are the CPU registers; at the
base is a very small number of secondary storage devices such as
solid-state or magnetic disks.  The hierarchy has two to four
intermediate layers: main memory (typically DRAM), and one to three
levels of CPU cache (typically SRAM).  The memory at the top of the
hierarchy is very fast, with a small blocksize: a word in a CPU
register typically holds 4 to 8 bytes.  The memory at the bottom of
the hierarchy is very slow, and it has a very large blocksize to
enable a generally-appropriate tradeoff of bandwidth for latency.  If
the blocksize of a disk transfer is too small, then any random or
linear access would deliver only a few bytes of useful data, and the
latency on this access might be millions of times larger than the CPU
cycle time -- so a computation that is bottlenecked at this level will
proceed at a rate of only a few bytes per millions of CPU cycles.
However if the blocksize is too large, then only a tiny fraction of a
randomly-accessed block will be useful.  As a general rule of thumb,
the blocksize of a layer of memory is the square root of its capacity.
The capacity $S_i$ of the $i$-th layer also seems to be a power
function (perhaps $S_i = (S_{i-1})^{1.4}$) of the capacity of the
layer immediately above it, with the random-access latency $L_i$ of
each level also growing as a power function (perhaps $L_i =
(L_{i-1})^{0.6}$) \cite{Thomborson1998}.

We can not predict absolute performance from a general model such as
the one above.  However we can predict relative performance, to an
accuracy of perhaps a factor of 10.  We do not believe it is feasible
to devise a general model that is more accurate than this, given the
diversity of contemporary computing platforms.  Indeed, we believe
that estimating relative performance within a factor of 10 in a
general model is a very challenging technical problem, even when the
computational workload is limited to the evaluation of Boolean
circuits.  On a general workload, a computation may be bottlenecked in
many different ways: by CPU instruction bandwidth, by CPU instruction
latency, by memory latency, by memory bandwidth, by latency or bandwidth
of interprocessor communication, or by power consumption (for heat-
or battery-limited computations).  We tentatively identify memory
latency as the most critical constraint, on most platforms, when 
they are evaluating large Boolean circuits.  Memory latency
bottlenecks arise when the working set is overly large.

We can measure the speed of a circuit evaluation in gate-evaluations
per second.  We expect our experimentation to confirm that this rate
is nearly constant, after a brief startup transient -- if the program
width (and thus the working set of its evaluator) does not vary
significantly by level.  We are moderately confident that the
obfuscated functions constructed under SafeWare will conform to this
expectation.  However, if constructions of obfuscated functions are
compositions of moderately-wide functions with very wide functions,
then (in subsequent work) we will adjust our programming language to
accomodate series-parallel functional compositions with declared
widths on each subcircuit.\footnote{The COPY operation in version 0x01
  of BPW will allow a width-$w$ evaluation of series-parallel
  compositions of arbitrary width-$w$ subfunctions, subject to the
  constraints that no subfunction has more than $\sqrt{w}$ outputs and
  that no more than $\sqrt{w}$ subcircuits are evaluated in parallel.}

Evaluation rates will almost surely be nondecreasing in the size of
the working set, but there will be very significant nonlinearities in
this relationship whenever the working set is almost equal to the
capacity of a layer of memory.  A fully-accurate timing model would be
parameterised on the threshold values of $w$ which (for a particular
evaluation method and a particular platform) are likely to cause this
evaluation method to become memory-bottlenecked at that level.  A
generally-valid timing model cannot have any platform-specific
parameters, so we restrict our attention to cost functions of the
form $nw^{\alpha}$.

Earlier in this section, we performed two platform-specific
estimations which suggested a factor of 40 difference between the
per-gate-evaluation time for a circuit which is too wide to be
efficiently evaluated, in comparison to a circuit which is narrow
enough to be efficiently evaluated.  The range of interest in circuit
width is 50 to 500000 -- a factor of 10000.  A factor of 40 difference
is, very roughly, a square root of this range; so we have seized on
the square root as a convenient exponent ($\alpha = 0.5$) for the
effect of circuit width $w$ on gate-evaluation rate, on any given
platform, for any given family of circuits.

Our proposed cost metric is thus $n\sqrt w$.  We may revise the
exponent on $w$, if experiments (such as the ones described in Section
4) on contemporary platforms of interest suggest that such revision
would be appropriate.  However we see very little chance that the
best-fit exponent $\alpha$, as determined by experimentation, will be
below $0.4$ or above $0.7$, except perhaps for long-running
computations on battery-powered platforms where (for theoretical
reasons \cite{Thomborson1998}) we would expect to observe
power-limited computations with runtimes proportional to $nw$.

\section{File Format}

In this section, we briefly sketch an efficient format for describing
large Boolean circuits of bounded width.  These files have the
extension .BPW, as an acronym for ``bounded program width'', so the
corresponding ``magic bytes'' must appear first in their header: 0x42,
0x50, 0x57.  The fourth byte is a version number.  Version 0x01 is
described in this article.

Four 8-byte integer parameters appear next in the file: $w$, $n$, $a$,
$b$.  The first parameter is the declared width $w$ of the circuit, as
represented in this BPW program.  The next parameter is the declared
number of gates $n$ in the circuit.  The third parameter is its number
of Boolean inputs.  Runtime arguments to the evaluation function would
supply these inputs, in order to determine the values of the circuit's
$b$ (the fourth parameter) Boolean outputs.

We require $a \leq w^2$ and $b \leq w^2$.  These may seem very
unnatural restrictions to a circuit-complexity theorist, because a
width-$w$ circuit of $n$ 3-input gates could naturally be allowed to
have up to $2w$ external inputs per level, with different inputs being
accepted on each of $\Omega(n/w)$ levels.  Such a circuit might
produce $n$ output bits.  However our emphasis is not on theoretical
elegance, but is instead on representing circuits so that they can be
evaluated efficiently on a contemporary computer system such as a
handheld device.

The body of a BPW file is a sequence of $n$ gate-descriptors.  Syntax
errors are clearly possible, for example the body of a BPW file may
not have exactly $n$ gate-descriptors.  A formal syntax for BPW is
outside of the scope of this article -- because our intent is to
sketch the initial (pre-release) version of this language in
sufficient detail that its design can be discussed, in a workshop
setting, prior to the finalisation of a first production version.

Each gate-descriptor starts with a 4-bit nibble encoding its type,
with the following possibilities (enumerated from 0 to 0xF): NOT,
AND2, OR2, NAND2, NOR2, XNOR2, AND3, OR3, NAND3, NOR3,
XOR3, XNOR3, MUX3, COPY, undefined.  Note that gate type 0xF is
undefined in version 0x01 of the BPW format.  Future formats
may use 0xF as a prefix for multiple-nibble gate-type descriptors.

Version 0x01 of BPW has one type of 1-input gate, six types of 2-input
gates, seven types of 3-input gates, and a three-input 'COPY'
pseudogate which is used for extended-length IO operations as well as
to represent width-$w$ parallel compositions while maintaining
locality in the state vector of the evaluation process.  We take the
(positive-logic) convention that 0 encodes a FALSE value and 1 encodes
a TRUE value.  The third input of MUX3 controls which of its first two
inputs should be copied onto its output, with a control of 0 selecting
the first input of this 2-input multiplexor.  The logic function of
every other gate type should be clear from its mnemonic.  We will
explain the semantics of the COPY pseudogate, immediately after
discussing the detailed semantics of logic-gate evaluation.

Gate input-specifiers are references to:
\begin{itemize}
\item any of $w$ external inputs (indexed as 0 to $w-1$),
\item any bit in a length-$w$ circular queue of results from
  prior gate-evaluations (indexed as $w$ to $2w-1$),
\item any (single-bit) result of an evaluation of a gate on the
  previous level of gates.  These results are stored in a length-$2w$
  circular queue, indexed as $2w$ to $4w-1$ in gate-descriptors; and
  these registers are locked against reads during the virtual-machine
  cycle in which they are being written.  Gate evaluations are done in
  a $w$-way parallel fashion, so only $w$ of these bit-registers are
  available as inputs while the other $w$ are being updated.
\end{itemize}

Summarising the above, input-specifiers are indexes into the register
file of a virtual machine with $4w$ bits of storage.  The virtual
machine state also has an instruction pointer (into the input stream
of the BPW file), and two register-pointers (of length $\lceil \lg 2w
\rceil$) which maintain the states of the two circular queues.

The reservations on bit-registers give BPW the flavour of a VLIW
instruction set, in which it is the programmer's responsibility to
schedule operations in order to acheive $w$-way parallelism.  The
resulting ``forbidden'' values of input-specifiers will be annoying to
human programmers, but will not be problematic for compilers.  We
expect circuit theorists to have little difficulty with this
representation, for (as explained previously) the theoretical notion
of circuit-width is involves the assignment of gates to levels, with
at most $w$ gates per level, under the restriction that gate inputs
are connected either to external inputs or to outputs from the
previous level.  If (as required in BPW) we have exactly $w$ gates per
level (after encoding nops \emph{e.g.} as OR2 gates with two inputs
connected to the output of the first gate on the preceding level), and
if we index the gates on even-numbered levels as $2w..3w-1$ and on
odd-numbered levels as $3w..4w-1$, then we have accurately specified
the interlayer connections in a BPW program.

The first $w$ external inputs are loaded into bit-registers 0 through
$w-1$ prior to the evaluation of the first gate-description by the
virtual machine.

The semantics of the COPY pseudo-gate is, we think, best explained by
its motivation: to mimic a cache fault which has been well-predicted
by a programmer.

Prior to the first gate-evaluation, the virtual machine's main memory
is initialised so that the $a$ external inputs to the circuit are
available in the first $\lceil a/w \rceil$ words of $w$ bits each,
in a randomly-accessible I/O space.  When gate-evaluations are
completed by the virtual machine, they are written into main memory at
successively higher addresses starting from address 0.  These writes
are naturally expressed as $w$-bit words, with each word representing
the evaluated outputs from one level of gates in a width-$w$ circuit.

The COPY pseudo-gate is a read-operation from the circuit inputs or
the main memory of the virtual machine, into its registers.  The first
operand of COPY is an input-specifier which identifies the relevant
word to be read.  The second operand specifies the number of bits to
be extracted from this word, and the third operand specifies the
starting bit-position within the retrieved word.  Destination
registers are specified implicitly: extended-input values are written
into the circular-queue of input registers (indexed as $0$ to $w-1$ by
input-specifiers), and prior-evaluation results are written into the
circular-queue of prior-result registers (indexed as $w$ to $2w-1$ by
input-specifiers).

If the first operand of a COPY is in the range $0..w-1$, then it is an
extended-input read.  For example COPY(1, 1, 0) is a read of the
$w$-th input bit.  The result is written into the next available
bit-register (indexed by virtual machine pointer PI) of the circular
queue $0..w-1$.  After every COPY, this register is incremented: PI +=
$1 \bmod w$.  Initially PI = 0.  Every BPW program starts with an
implicit COPY(0, $w$, 0) -- so that the first $w$ circuit inputs are
available for immediate use in gate-evaluations.

If the first operand of a COPY is in the range $w..2w-1$, then it is a
prior-result read. The target address of every such read is relative
to the virtual machine's prior-result pointer $PR$, whose value is
initially 0.  The prior-result pointer of the machine is incremented
by one after each gate-evaluation: $\mbox{\rm PR} = (\mbox{\rm PR} +
1) \bmod 2w) + w$.  The output of the $i$-th gate on the $j$-th
preceding level is accessible as COPY($j$, 1, $i$).

The latency of a COPY operation is $\sqrt w$, as measured in VLIW
instruction times; it is $w \sqrt w$, if measured in gate-evaluation
instructions.  This latency is enforced by syntactic restrictions on
BPW version 0x01.  We call such programs BPW1 for convenience.
Subsequent versions of BPW may have different restrictions.
A BPW1 program is invalid if
\begin{itemize}
\item it contains more than one COPY operation per $w$ instructions, or
\item if the result of a COPY by a BPW instruction at level $i$ is referenced by
  a gate-evaluation instruction at level $j < i + \sqrt{w}$.
\end{itemize}
Levels are defined by counting the non-COPY instructions in a program,
with level 0 being the initial level, and the level counter being
incremented after every $w$ non-COPY instructions.  As previously
indicated, a gate-evaluation has latency 1, that is, its result is
unavailable to gate-evaluations in the same level but is available
to gate-evaluations in the subsequent level.

If the first operand of a COPY is in the range $2w..3w-1$, then its
semantics are undefined in version 0x01 of BPW.  In future versions,
BPW semantics may be extended to allow more external inputs (perhaps
up to $w^4$) and more prior-results (perhaps up to $w^4$) to be
accessible, with an appropriately-high latency (perhaps $w^2$).

In version 1 of BPW, we emphasise programming convenience (and CPU
cycle-timing) over file compression; so we nibble-align all of the
input-specifiers (rather than bit-aligning them for optimal
compression).  The length, in nibbles, of each input-specifier is
$\lceil \lg{4w}/2 \rceil$.  For example, a width-50 circuit requires 1
byte (2 nibbles) for each input-specifier.  If it were composed
entirely of 2-input gates, then each gate (with its 1-nibble type) is
encoded in 2.5 bytes.  A disorganised circuit with a million gates
could be declared to have width 500,000; so it would require 5 nibbles
for each input-specifier, for a total of 5.5 bytes per gate.

\section{Experimental Designs}

In this section, we briefly sketch some experimentation which would
validate (or invalidate) our cost function and our BPW language
design.

\subsection{Workload}
The experimental workload consists of two types of BPW programs at
varying $n$ (size), $w$ (width), and $d$ (density of COPY
operations), for all 
{\setlength\arraycolsep{2pt}
\begin{eqnarray}
  w & = & \{5, 10, 50, 100, 500, 1000, 5000, 10000, \nonumber\\
  & & \ \ 50000, 100000, 500000\}, \\
  n & = & \{10^6, 10^7, 10^8, 10^9 \}, \\
  d & = & \{ {1 \over w}, {1 \over {2w}}, {1 \over {4w}}, 
    ..., {1 \over {2^{\lfloor \lg (n/w)\rfloor}w}} \}.
\end{eqnarray}}.

The first type of program implements a randomly-chosen function with
$k = \min(w, 50)$ inputs and $k$ outputs.  The second type of program
is a lightly obfuscated password recogniser, for the very
insecurely-chosen $k$-bit password 0x1555...555.  The output of the
password recogniser is 1 if the input matches the password, 0
otherwise.

The first type of program consists of a sequence of $nd/(1+d)$
repetitions of the following subcircuit: $1/d$ randomly-generated
NAND2 gates, a randomly-generated COPY pseudogate-specifier.  The
input-specifiers on the NAND2 gates are generated from i.u.d.
variates on $\{ 0, 1, ..., 4w-1\}$, discarding any generated values
which are invalid due to the uninitialised or unavailable
bit-registers at this point in the BPW program.  The first
input-specifier on each COPY pseudogate is i.u.d. on the
currently-valid values in $\{ w, w+1, ..., 3w-1\}$.  The second
input-specifier of each COPY pseudogate is $\lfloor \sqrt w / 2
\rfloor$, the number of bit-registers to be written.  When such COPY
pseudogates appear in a BPW program at density $d = 1/w$, then (due to
the latency of COPY operations) about half of the prior-results cache
is being updated at any time during the program execution.  The other
half of the prior-results cache may be referenced by input-specifiers.

The password-recogniser should compute $w$ copies of its output bit in
its last $\lceil \lg \min(w, 50) \rceil$ levels, using XOR2 and XNOR2
gates on the first of these levels and using AND2 gates for the
remaining levels.  Note that the $i$-th bit of the secret password is
encoded in the type (XOR2 or XNOR2) of the gate which receives two
copies of the $i$-th input bit. The input bits are permuted in the
intermediate levels of the password-recogniser.  Each intermediate
level is composed of $w$ NOT gates with input-specifiers which
(collectively) define a random permutation on $w$ elements -- so that
each intermediate level is a very weak obfuscation (by bit-scrambling)
of the input.  The first level of the password-recogniser is also
composed of NOT gates, with the $i$-th gate having input-specifier $i
\bmod \min(w, 50)$.

\subsection{Systems under test} A BPW execution environment should be
set up on a mid-spec smartphone (such as a Samsung Galaxy III S), a
mid-spec laptop computer, a mid-spec desktop computer (using its CPU
as the function evaluator), and a mid-spec desktop computer using its
CUDA-enabled GPU as the function evaluator.

\subsection{Primary measurements} The experimenter should measure the
runtime and total energy consumption of the function evaluator,
exclusive of loading the circuit description into the primary memory
of the computing platform.  The function evaluator should be coded in
three different ways:
\begin{enumerate}
\item For ease of programming, with the 4$w$ bits of evaluation state
($w$ inputs, $w$ prior subcircuit outputs, $w$ current subcircuit outputs,
$w$ copied outputs from a prior subcircuit in a parallel composition)
stored in a single machine-addressible array of 4$w$ bytes or words;
\item For memory latency, with the 4$w$ bits of evaluation state
in a bit-packed array;
\item To avoid CPU and GPU bottlenecks, if the resourcing of the
  experimental team permits this: BPW programs should be compiled into
  machine code that is well-optimised for each platform.  We expect
  such compilations to avoid CPU bottlenecks on any platform, except
  for very small $w$; whereas the byte-by-byte interpretations of the
  first two codings may introduce CPU bottlenecks for $w$ up to 500,
  and GPU bottlenecks may be unavoidable unless the code is compiled.
\end{enumerate}

\subsection{Secondary measurements} The experimenter should collect a
timeseries, at 10 msec intervals, of the temperature and cycle rate of
the CPU or GPU.  These timeseries should be annoted, to indicate the
start-time, stop-time, and identity (program type, $n$, $w$, $d$,
type of evaluation method) of each BPW evaluation in the experimental
sequence.  The platform under test should be allowed to cool down to a
baseline temperature before starting another evaluation.

\subsection{Hypotheses under test} 

\begin{enumerate}

\item On each platform, for both program types, and for each of its
  available evaluation routines: confirm that the runtime for each $w$
  is linear in $n$, and is nondecreasing in $w$, with a factor of
  about $\sqrt{500000/50} = 320$ separating the runtime curve for
  $w=50$ from the runtime curve for $w = 500000$.  Fail to accept the
  hypothesis if the computed separation in runtime for small and large
  $w$ is either less than 32 or greater than 3200, for any platform,
  circuit type, or evaluation routine.

\item Compute a best-fit value of a speedup ratio $R$ for each
  platform, function type, and evaluation type, where $R$ is the average
  speedup (over all feasible $n$) for the evaluation of a width-50
  circuit as compared to the evaluation of a width-500000 circuit.
  Confirm that the best-fit value for $R$ is not significantly
  affected by platform, function, or evaluation method.

\item Compute a ratio of the total energy consumption of each
  computation to the value $nw$, this formula being a theoretical
  prediction of the energy consumption of a memory-limited computation
  on a computational device that is optimised for energy efficiency
  \cite{Thomborson1998}.  Confirm that, for $w=500000$, this
  ratio (a measure of peta-reference-bytes per watt-hour) is within a
  factor of 10 across all platforms.

\item If experimental resources permit, perform additional
  experimentation on larger $n$ with very large $w$ to confirm that a
  power bottleneck is possible \emph{i.e.}  that the CPU or GPU speed
  has been throttled to avoid overheating.  If power bottlenecks are
  commonly observed within the range of practical interest, then the
  cost function $nw$ should be considered as a possible replacement
  for the proposed $n\sqrt{w}$.

\end{enumerate}

\section{Summary and Discussion}

We have discussed the promise of \emph{indistiguishability
  obfuscation} (iO) as a technique for obfuscating programs, we have
proposed a method for estimating the time required to evaluate an
obfuscated function, and we have proposed an experimental method for
validating our proposed estimation method.

At the time of writing, iO is a very promising theory that has not yet
been reduced to practice.  There are no published methods for
producing feasibly-computable iO circuits for any functions of
practical importance, such as the recognition of a 50-bit password.

Our BPW language is directly comparable to Barrington's a $w$-BP
language \cite{Barrington1989}.  Regrettably, Barrington's programs do
not have easily-predictable performance on real-world computer
systems, because their unrestricted references to inputs may result in
IO bottlenecks.  In Barrington's circuit-theoretical context, any
charge for access to inputs ``would lead to a class far too restricted
to be interesting'' \cite{Barrington1989}.

In a possible variant of BPW (or $w$-BP) which models online
computations, an input stream of unbounded length may be presented on
a one-way read-only tape.  If such IO streams ever become a promising
line of theoretical enquiry for functional obfuscation, then a future
version of BPW should allow streamed-IO -- at some blocksize and
latency that has been experimentally determined to be feasibly
achievable, on a wide variety of contemporary mass-market computing
platforms in typical networking environments.

Barrington has proved that any language recognised by an $\mbox{\rm
  NC}^1$ circuit can be recognised by 5-PBP, that is, by a restricted
5-BP in which all of the $w$-maps are permutations.  Each instruction
in a 5-BPB could be implemented in 3$w$ BPW instructions in a
computation of declared width $4w$, if the length of the input is
bounded by $w$.  This construction may be devoid of practical
relevance, because of its very restrictive input bound, and because
Barrington's 5-PBP program is exponential in the depth of the
$\mbox{\rm NC}^1$ circuit.

We suspect that partial evaluations will be important in many
applications of functional obfuscation.  This could be handled via the
syntax of the call to the functional evaluator, whereby the output of
the partial evaluation is a new BPW program which describes an
efficiently-computable projection of the original program with a
correspondingly-reduced set of inputs. 

\bibliographystyle{IEEEtran}  

\bibliography{IEEEabrv,benchfo} 

\begin{thebibliography}{10}
\providecommand{\url}[1]{#1}
\csname url@samestyle\endcsname
\providecommand{\newblock}{\relax}
\providecommand{\bibinfo}[2]{#2}
\providecommand{\BIBentrySTDinterwordspacing}{\spaceskip=0pt\relax}
\providecommand{\BIBentryALTinterwordstretchfactor}{4}
\providecommand{\BIBentryALTinterwordspacing}{\spaceskip=\fontdimen2\font plus
\BIBentryALTinterwordstretchfactor\fontdimen3\font minus
  \fontdimen4\font\relax}
\providecommand{\BIBforeignlanguage}[2]{{%
\expandafter\ifx\csname l@#1\endcsname\relax
\typeout{** WARNING: IEEEtran.bst: No hyphenation pattern has been}%
\typeout{** loaded for the language `#1'. Using the pattern for}%
\typeout{** the default language instead.}%
\else
\language=\csname l@#1\endcsname
\fi
#2}}
\providecommand{\BIBdecl}{\relax}
\BIBdecl

\bibitem{Garg2013}
S.~Garg, C.~Gentry, S.~Halevi, M.~Raykova, A.~Sahai, and B.~Waters, ``Candidate
  indistinguishability obfuscation and functional encryption for all
  circuits,'' in \emph{54th Annual Symposium on Foundations of Computer Science
  (FOCS)}, Oct 2013, pp. 40--49.

\bibitem{SafeWare2014}
{Defense Advanced Research Projects Agency}, ``Broad agency announcement:
  {SafeWare} {DARPA}-{BAA}-14-65,'' Sep. 2014.

\bibitem{Collberg2002}
C.~Collberg and C.~Thomborson, ``Watermarking, tamper-proofing, and obfuscation
  - {T}ools for software protection,'' \emph{IEEE Transactions on Software
  Engineering}, vol.~28, no.~8, pp. 735--746, Aug 2002.

\bibitem{LaDue1997}
\BIBentryALTinterwordspacing
M.~LaDue, ``The {M}aginot license: {F}ailed approaches to licensing {J}ava
  software over the {I}nternet,'' copyright 1997, available 27 Jan. 1998.
  [Online]. Available: \url{http://www.rstcorp.com/hostile-applets/Maginot/}
\BIBentrySTDinterwordspacing

\bibitem{VHDL2008}
\emph{{IEEE} Standard {VHDL} Language Reference Manual}, {IEEE} Computer
  Society, 2008.

\bibitem{Bogetoft2009}
\BIBentryALTinterwordspacing
P.~Bogetoft, D.~Christensen, I.~Damg{\aa}rd, M.~Geisler, T.~Jakobsen,
  M.~Kr{\o}igaard, J.~Nielsen, J.~Nielsen, K.~Nielsen, J.~Pagter,
  M.~Schwartzbach, and T.~Toft, ``\BIBforeignlanguage{English}{Secure
  multiparty computation goes live},'' in
  \emph{\BIBforeignlanguage{English}{Financial Cryptography and Data
  Security}}, ser. Lecture Notes in Computer Science, R.~Dingledine and
  P.~Golle, Eds.\hskip 1em plus 0.5em minus 0.4em\relax Springer Berlin
  Heidelberg, 2009, vol. 5628, pp. 325--343. [Online]. Available:
  \url{http://dx.doi.org/10.1007/978-3-642-03549-4\_20}
\BIBentrySTDinterwordspacing

\bibitem{Pippenger1979}
N.~Pippenger, ``On simultaneous resource bounds,'' in \emph{20th Annual
  Symposium on Foundations of Computer Science (FOCS)}, Oct 1979, pp. 307--311.

\bibitem{Dymond1989}
\BIBentryALTinterwordspacing
P.~W. Dymond and S.~A. Cook, ``Complexity theory of parallel time and
  hardware,'' \emph{Information and Computation}, vol.~80, no.~3, pp. 205--226,
  1989. [Online]. Available:
  \url{http://dx.doi.org/10.1016/0890-5401(89)90009-6}
\BIBentrySTDinterwordspacing

\bibitem{Allender1990}
E.~Allender and C.~Wilson, ``Width-bounded reducibility and binary search over
  complexity classes,'' in \emph{Proc. 5th Annual Structure in Complexity
  Theory Conference}, July 1990, pp. 122--129.

\bibitem{Yoo2013}
\BIBentryALTinterwordspacing
R.~M. Yoo, C.~J. Hughes, K.~Lai, and R.~Rajwar, ``Performance evaluation of
  {Intel} transactional synchronization extensions for high-performance
  computing,'' in \emph{Proceedings of the International Conference on High
  Performance Computing, Networking, Storage and Analysis}, ser. SC '13.\hskip
  1em plus 0.5em minus 0.4em\relax New York, NY, USA: ACM, 2013, pp.
  19:1--19:11. [Online]. Available:
  \url{http://dx.doi.org/10.1145/2503210.2503232}
\BIBentrySTDinterwordspacing

\bibitem{Fermi2009}
\BIBentryALTinterwordspacing
\emph{{NVIDIA}'s Next Generation {CUDA} Compute Architecture: {Fermi}}, NVIDIA
  Corporation, 2009, v1.1. [Online]. Available:
  \url{http://international.download.nvidia.com/pdf/kepler/NVIDIA-Kepler-GK110-GK210-Architecture-Whitepaper.pdf}
\BIBentrySTDinterwordspacing

\bibitem{Thomborson1998}
\BIBentryALTinterwordspacing
C.~D. Thomborson, ``The economics of large-memory computations,''
  \emph{Information Processing Letters}, vol.~66, no.~5, pp. 263--268, 1998.
  [Online]. Available: \url{http://dx.doi.org/10.1016/S0020-0190(98)00063-5}
\BIBentrySTDinterwordspacing

\bibitem{Barrington1989}
\BIBentryALTinterwordspacing
D.~A. Barrington, ``Bounded-width polynomial-size branching programs recognize
  exactly those languages in {$\mbox{\rm NC}^1$},'' \emph{Journal of Computer
  and System Sciences}, vol.~38, no.~1, pp. 150 -- 164, 1989. [Online].
  Available: \url{http://dx.doi.org/10.1016/0022-0000(89)90037-8}
\BIBentrySTDinterwordspacing

\bibitem{SavageMHG}
J.~E. Savage, \emph{Models of Computation: Exploring the Power of
  Computing}.\hskip 1em plus 0.5em minus 0.4em\relax Addison-Wesley, 1997, ch.
  11, Memory-Hierarchy Tradeoffs.

\end{thebibliography}

\end{document}